# Probability Prediction based Reliable Opportunistic (PRO) Routing Algorithm for VANETs

Ning Li, *Student Member*, *IEEE*, Jose-Fernan Martinez-Ortega, Vicente Hernandez Diaz, Jose Antonio Sanchez Fernandez

*Abstract*–**In the Vehicular ad hoc networks (VANETs), due to the high mobility of vehicles, the network parameters change frequently and the information which the sender maintains may outdate when it wants to transmit data packet to the receiver, so for improving the routing effective, we propose the probability prediction based reliable (PRO) opportunistic routing for VANETs. The PRO routing algorithm can predict the variation of Signal to Interference plus Noise Ratio (SINR) and packet queue length (PQL) in the receiver. The prediction results are used to determine the utility of each relaying vehicle in the candidate set. The calculation of the vehicle utility is weight based algorithm and the weights are the variances of SINR and PQL of the candidate relaying vehicles. The relaying priority of each relaying vehicle is determined by the value of the utility. By these innovations, the PRO can achieve better routing performance (such as the packet delivery ratio, the end-to-end delay, and the network throughput) than the SRPE, ExOR (street-centric), and GPSR routing algorithms.**

## I. INTRODUCTION

Vehicular ad hoc network (VANETs) is a kind of network which combines the wireless communication with the vehicles to enable the vehicles to communicate with each other [1][2]. Due to the specific characteristics of VANETs, the VANETs are quite different with the traditional mobile ad hoc networks (MANETs). For instance, the speed of vehicles in VANETs is much higher than that in MANETs; the moving directions of vehicles in VANETs are limited by the urban streets; higher probability of network partition in VANETs than that in MANETs due to the traffic light [3]; due to the different structures of the streets (for instance, one-/two-way street, two-/four-lanes street), the network topologies are quite different with different streets [4], which is called topology diversity. Therefore, the routing algorithms in VANETs are different with that in traditional MANETs. The routing algorithms which are effective in MANETs may have poor performance in VANETs.

There are two routing strategies for the VANETs: deterministic routing and opportunistic routing [5]. In deterministic routing, the sender sends data packet to one neighbor vehicle which is chosen based on the optimal algorithms. In opportunistic routing, the sender sends the data packet to a set of relaying vehicles rather than only one relaying vehicle to improve the packet delivery ratio between sender and receiver. In this paper, we mainly focus on the opportunistic routing.

### A. Motivation

The main advantage of opportunistic routing compared with the deterministic routing is that it can improve the packet delivery ratio greatly [6][7]. In the opportunistic routing, one of the crucial parameters which can affect the packet delivery ratio between the sender and the receiver is the Signal to Interference plus Noise Ratio (SINR). If the receiver can receive the data packet that transmitted from the sender correctly, the SINR at the receiver must larger than the receiving threshold [8][9][10]. In the previous work, the SINR in wireless network has been investigated in-depth and many high quality routing algorithms have been proposed, such as [9], [10], [11], and [12]. However, in these algorithms, the calculation of the SINR is not sufficient to reflect the dynamic of the network, especially in the VANETs. In VANETs, due to the high mobility of vehicles, the network parameters change frequently and the information which the sender maintains may outdate when it wants to transmit data packet to the receiver [13]. So for improving the routing performance, the routing algorithms should be able to predict the variation of the network parameters (i.e. the SINR) before data packet transmission. In [13], [14], and [15], the link availability prediction has been investigated; during the routing process, the candidate relaying node which the predicted link availability is higher has higher relaying priority. However, how to predict the variation of the SINR in VANETs has not been investigated in the previous works; so in this paper, we will present the research on this issue in detail.

Moreover, not only the SINR, but the packet queue length (PQL) in the buffer of the vehicle also has great effection on the packet delivery ratio between the sender and the receiver. As introduced in [13], the data traffic may be aggregated at some vehicles through improper routing, which incurs long PQL (i.e. long one-hop delay), even worse may induce buffer overflow, leading to packet drops at network layer. In [13], the authors investigate the issue about selecting the proper relaying routing to improve the routing performance; however, the effection of the PQL on the relaying node selection has not been investigated. For instance, assuming that the SINR of the receiver is larger than the receiving threshold, but the residual buffer is not large enough to store the data packet, the data packet will be dropped. This means the packet delivery ratio reduces and the transmission delay increases. So the PQL of the receiver should also be taken into account during the relaying node selection. Similar to SINR, for improving the routing performance, the routing algorithm should also be able to predict the variation of the PQL, which is not investigated in the previous works. This is the second research item of this paper.

### B. Main contributions

Motivated by the issues introduced above, in this paper, we propose the probability prediction based reliable opportunistic routing (PRO) algorithm for VANETs, which can predict the variation of the network parameters (the SINR and the PQL).

The main contributions of this paper can be summarized as:

1. We propose two probability prediction algorithms: 1) probability prediction algorithm of SINR, which is used to

Ning Li, Jose-Fernan Martinez-Ortega, Vecente Hernandez Diaz, and Jose Antonio Sanchez Fernandez is with the Universidad Politenica de Madrid, Madrid, Spain.
E-mail: {li.ning, jf.martinez, vicente.hernandez, j.sanchez}@upm.es.

calculate the probability that the SINR of the receiver is larger than the receiving threshold after $\Delta t$ ; in this algorithm, both the effection of the number of neighbors and their distances to the receiver are taken into account; 2) probability prediction algorithm of PQL, which is used to predict the probability that the receiver's PQL is smaller than the maximum allowed value after $\Delta t$ ; in this algorithm, the nodes move-in and move-out the transmission area of the receiver are considered;

2. Based on the network parameter prediction algorithms introduced above, we propose the weight based candidate set selection algorithm for VANETs; in this algorithm, the utilities of the relaying vehicles in the candidate set are calculated based on the predicted SINR and PQL; the vehicle which has higher quality performance on both the SINR and PQL has higher utility; the number of vehicles in the candidate set is determined by the packet delivery ratio between the sender vehicle and the candidate set.

The remaining of this paper is organized as follows: in Section II, we review the related works in recent years; Section III defines the network model used in this paper; in Section IV, we introduce the probability prediction algorithms of SINR and packet queue length (PQL); Section V introduces the principle of the PRO routing algorithm in detail; in Section VI, the routing performance of the PRO algorithm, the SRPE algorithm, the GPSR algorithm, and the ExOR (street-centric) algorithm are evaluated and compared; Section VII concludes our work in this paper.

## II. Related Works

There are many routing algorithms has been proposed in recent years to address different issues of VANETs. In [16], considering the network densities are different to different networks, the authors propose two routing algorithms for sparse network and dense network, respectively. The algorithms identify the network density by using the number of two-hop neighbors. If the network is dense, then the routing decision is based on the neighbors' position; otherwise, both the position and the moving directions are used. For the 1-D two-way linear VANETs, the authors in [17] propose an epidemic routing; moreover, for investigating the performance of this routing algorithm, a finite-state Markov chain based stochastic model has been developed. In VANETs, one of the important issues is the quality of service for the video on demand (VOD) session. For solving this issue, in [18], the authors proposed a simplex VOD transmission algorithm for urban environment. In this algorithm, a set of independent routes are founded between the source vehicle and destination vehicle before the data packet transmission. The number of routes is decided by the volume of video and the lifetime of each route. For selecting the best connected route, a closed form equation which is used to estimate the connectivity probability of route has proposed. A concept called micro-topology (MT) is proposed in [19], which includes the vehicles and the wireless links in the street. During the relaying node selection, the MT rather than the single vehicle will be chosen as the next hop relaying unit. In [20], based on the stochastic analysis, the authors investigate the impact of cluster instability on the generic routing overhead. In this algorithm, the time variation of the cluster structure (include the cluster membership change and the cluster overlap state change rate) is taken into account. In [21], for improving the reliability of VANETs, the authors introduce the any-path into the routing design and propose a long lifetime any-path algorithm to improve the link stability of the VANETs. Similar to [21], in [22], the authors propose the PFQ-AODV algorithm to improve the reliability of the routing algorithm in VANETs. In PFQ-AODV, the fuzzy constraint Q-learning algorithm has been introduced into the AODV algorithm to improve the routing performance. The authors in [14] use the evolving theory to model the communication graph on a highway to improve the routing reliability. Based on this model, an evolving graph-based reliable routing scheme has been proposed. In [23], the authors propose Greedy Perimeter Stateless Routing (GPSR) for wireless datagram network. In GPSR, the node uses the information about the router's immediate neighbors to make greedy forwarding decision. In case a packet reaches a routing void, then the GPSR is recovered by routing around the perimeter of this region. Since there are so many routing algorithms for the VANETs, we can not introduce all of them in this paper, the more routing algorithms can be found in [15, 24-29].

In the traditional routing strategy, the packet delivery ratio is low since the source node chooses only one next hop relaying node; for improving the packet delivery ratio, the opportunistic routing has been proposed in [5]. In recent years, the opportunistic routing has been introduced into the VANETs to improving the packet delivery ratio. In [13], the authors propose a link availability probability prediction model and a new concept called the link correlation which is used to represents the influence of different link combinations. Based on these conclusions, a street-centric opportunistic routing protocol which based on the expected transmission cost over a multi-hop path has been proposed. Considering the degradation of delivery ratio, the authors in [30] introduce the opportunistic routing into the geographical source routing. In [31], for improving the packet delivery ratio and reducing the link breakage probability, first, the authors propose a hybrid approach to filter and prioritize the candidate set; then a flexible opportunistic forwarding strategy has been designed, in which the multiple neighbors of the sender has been taken into the local forwarding. In [32], considering the immoderate utilization of wireless fading channels could incur high distortion due to high probabilities of video package loss and damage, the authors take the interference into account and formulate the rate distortion model for live video streaming in VANETs. Based on this model, the authors propose the routing algorithm which can seek a balance between the distortion and delay. More opportunistic routing based algorithms for VANETs can be found in [33]-[37].

## III. Network Model

The network model used in this paper is shown in Fig. 1. In VANETs, the vehicles can only move along the streets. Each vehicle uses the same transmission power to communicate with other vehicles, i.e. the transmission rages of different vehicles are the same. At the intersections, only one road is unblocked, which can be found in Fig. 1. The velocity variation follows a truncated Gaussian distribution as that shown in [38]; moreover, in this paper, the *Wiener process* is utilized to model the movement of vehicles [39][40]. On each road, there are two moving directions. Moreover, the vehicles which locate at different intersections of the same road can not communicate with each other directly; for instance, as shown in Fig. 1, the vehicle $a$ and vehicle $b$ can not communicate with each other directly. In the network, two vehicles can communicate directly when there is a bi-directorial communication link between these two vehicles. The bi-directorial communication link means that two vehicles can communicate with each other without relaying by the third vehicle. For instance, if the vehicle $v_s$ can communicate with $v_d$ directly, then $d_{sr} \le r_s$ and $d_{sr} \le r_d$ ; $d_{sr}$ is the Euclidean distance between $v_s$ and $v_d$. The vehicles equip GPS devices and can acquire their location.

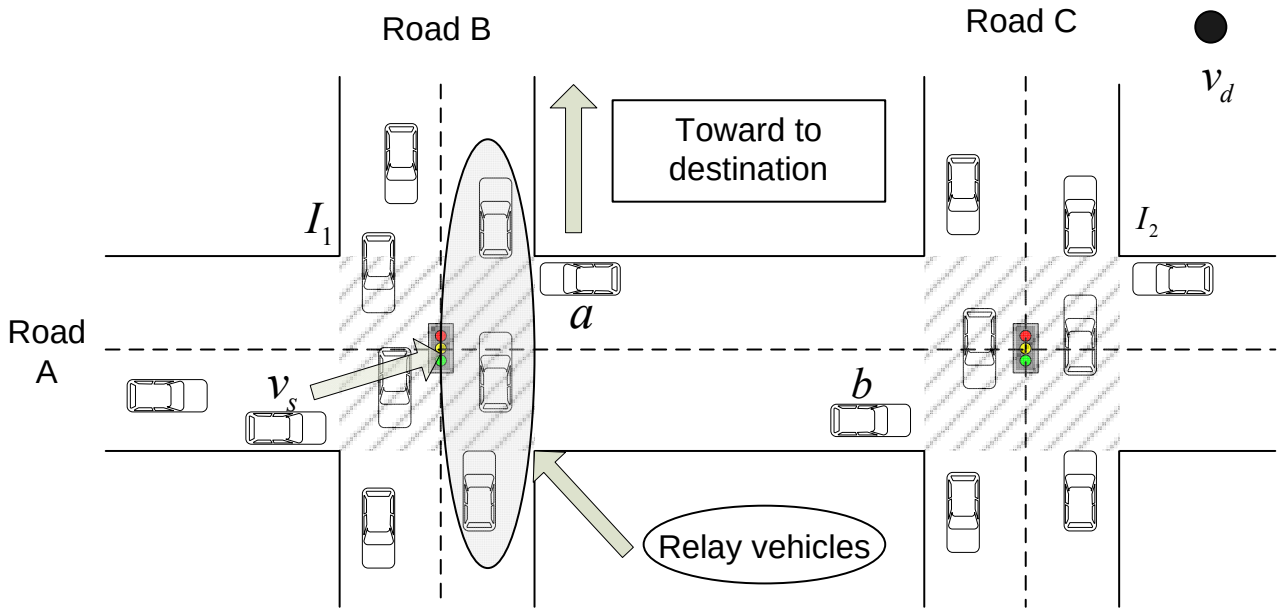

Fig. 1. Network model

For a transmission pair $\langle v_s, v_d \rangle$ , the sender $v_s$ can move close to or faraway from the destination $v_d$. In this paper, the sender vehicle $v_s$ moves close to the destination vehicle $v_d$ means that after $\Delta t$ , the Euclidean distance between $v_s$ and $v_d$ reduces. Based on this, the vehicle $v_s$ moves close to vehicle $v_d$ can be divided into two directions: horizontal direction and vertical direction. For instance, to the transmission pair $\langle v_s, v_d \rangle$ (the $v_d$ has been shown in Fig. 1), the moving directions of $v_s$ toward to $v_d$ are: 1) upward movement along Road *B* or Road *C*; 2) moving to the right along Road *A*.

## IV. PROBABILITY PREDICTION ALGORITHM

Based on the mobility model of the vehicles in VANETs [38][39][40], the sender vehicle can predict the parameter variation of its neighbor vehicles. The probability prediction algorithms of the distance variation and the link availability have been proposed in previous works, such as [13], [39], [40], [41], [42], so the main content in this section is to introduce the probability prediction algorithm of SINR and PQL which have not been investigated in the previous works.

In [13], [39], [40], and [41], the distance variation and link availability prediction algorithms have been proposed. For predicting the distances variation between the vehicle *v* and its neighbors, the moving pattern introduced in [38] is referenced in [13]. The velocity variation $\Delta v$ of vehicle *i* is calculated as:

$$\Delta v_{i,t_{12}} = v_{i,t_1} - v_{i,t_2} = \sigma_i \sqrt{t_2 - t_1} \qquad (1)$$

where $v_{i,t_1}$ and $v_{i,t_2}$ are the velocities of vehicle $v_i$ at $t_1$ and $t_2$, respectively; $\Delta v_{i,t_{12}}$ is the velocity changing of $v_i$ during $[t_1, t_2]$ ; $\sigma_i$ follows a standard Gaussian distribution. Additionally, $\Delta v$ has an independent increment under different time intervals. So the relative distance changing between vehicles $v_i$ and $v_j$ during $[t_1, t_2]$ can be expressed as:

$$\Delta d(v_i, v_j)_{\Delta t} = (v_{i,t_1} - v_{j,t_1} + \Delta v_{i,t_{12}} - \Delta v_{j,t_{12}}) \cdot \Delta t \qquad (2)$$

where $\Delta t = t_2 - t_1$ . The value of $\Delta d(v_i, v_j)_{\Delta t}$ relates to both the $v_{i,t_1}$ , $v_{j,t_1}$ , $\Delta v_{i,t_{12}}$ , and $\Delta v_{j,t_{12}}$ . According to different values of $v_{i,t_1}$ , $v_{j,t_1}$ , $\Delta v_{i,t_{12}}$ , and $\Delta v_{j,t_{12}}$ , the $\Delta d(v_i, v_j)_{\Delta t}$ could larger or smaller than 0, which means the vehicles could close to or far away from each other. In [13], the authors has proved that $\Delta v_{i,t_{12}}$ , $\Delta v_{j,t_{12}}$ , and $\Delta v_i - \Delta v_j$ are all independent variables and follow the zero-mean Gaussian distribution. Moreover, after $\Delta t$ , the probability that the vehicle $v_s$ and $v_r$ can communicate with each other can be calculated as:

$$\begin{aligned} P_{sr}^{d}(\Delta t) &= p\left\{\Delta d(v_s, v_r)_{\Delta t} < R - d_{sr,t_1}\right\} \\ &= F\left(\Delta v_{i,t_{12}} - \Delta v_{j,t_{12}} < \frac{R - d_{sr,t_1} - \left(v_{i,t_1} - v_{j,t_1}\right)\Delta t}{\Delta t}\right) \\ &= \int_{-\infty}^{R-d_{sr,t_1}} f\left(\Delta v_{i,t_{12}} - \Delta v_{j,t_{12}}\right) d\left(\Delta v_{i,t_{12}} - \Delta v_{j,t_{12}}\right) \end{aligned} \qquad (3)$$

where $\Delta v_i \sim N(0, \sigma_i^2 \Delta t)$ , $\Delta v_j \sim N(0, \sigma_j^2 \Delta t)$ , and $\Delta v_i - \Delta v_j \sim N(0, \sigma_i^2 \Delta t + \sigma_j^2 \Delta t)$ .

Based on the conclusions and assumptions introduced above, we propose the prediction algorithms of SINR and PQL in the following of this section.

### *A. Prediction algorithm of SINR*

The SINR at the receiver can be calculated according to the conclusions in [43]. When the vehicle $v_s$ sends data packet to vehicle $v_r$ , the interference of vehicle $v_r$ at $t_1$ can be expressed as:

$$I_{r,t_1} = N + \sum_{i=1}^{n_{int-v,t_1}} \xi(d_{ir,t_1}) = N + \sum_{i=1}^{n_{int-v,t_1}} \frac{P_t}{(d_{ir,t_1})^{\alpha}} \cdot \frac{G_t G_r \lambda^2}{(4\pi)^2 L} \qquad (4)$$

where $n_{int-v,t_1}$ is the number of interference vehicles at $t_1$, *N* is the additive white Gaussian noise (AWGN), $\alpha$ is the path loss exponent and $2 \le \alpha \le 5$ depends on the geometry of propagation environment [42]. Without lose of generality, let $G = \frac{G_t G_r \lambda^2}{(4\pi)^2 L}$ , therefore, (4) can be rewritten as:

$$I_{r,t_1} = N + G \sum_{i=1}^{n_{int-v,t_1}} \frac{P_t}{(d_{ir,t_1})^{\alpha}} \qquad (5)$$

If the vehicle $v_r$ can receive the data packet that transmitted from vehicle $v_s$ successfully, the SINR at node $v_r$ should satisfy the constraint as follows:

$$SINR_{r,t_1} = \frac{G \frac{P_t}{(d_{ir,t_1})^{\alpha}}}{N + G \sum_{i=1, i \ne s}^{n_{int-v,t_1}} \frac{P_t}{(d_{ir,t_1})^{\alpha}}} = \frac{(d_{sr,t_1})^{-\alpha}}{N + \sum_{i=1, i \ne s}^{n_{iint-v,t_1}} (d_{ir,t_1})^{-\alpha}} \ge \beta \qquad (6)$$

In (6), as the assumptions in Section III, all the nodes have the same transmission power $P_t$; $d_{sr,t_1}$ is the Euclidean distance between $v_s$ and $v_r$ at $t_1$; $d_{ir,t_1}$ ( $i = 1,2,...,m$ ) is the Euclidean distance between the receiver $v_r$ and its interference vehicles at $t_1$ (the interference vehicle is defined as the vehicle which the transmission range covers $v_r$ ); $\beta$ is the receiving threshold which can guarantee successful data packet decoding at the receiver.

According to the conclusions in [13] and (2), the distance variation between the sender vehicle $v_s$ and the receiver vehicle $v_r$ after $\Delta t$ can be calculated as:

$$d_{sr,\Delta t} = d_{sr,t_1} + \Delta d(v_s, v_r)_{\Delta t} \qquad (7)$$

The distance variation between the receiver $v_r$ and its interference vehicles after $\Delta t$ can be expressed as:

$$d_{ir,\Delta t} = d_{ir,t_1} + \Delta d(v_i, v_r)_{\Delta t},\ i = 1,2,...,m \quad (8)$$

As shown in (7), when the distance between the sender vehicle $v_s$ and the receiver vehicle $v_r$ is $d_{sr,t_1}$ at $t_1$, then after $\Delta t$, the distance between $v_s$ and $v_r$ will be $d_{sr,\Delta t}$. Additionally, since $\Delta v_s \sim N(0,\sigma_s^2 \Delta t)$, $\Delta v_r \sim N(0,\sigma_r^2 \Delta t)$, and $\Delta v_s - \Delta v_r \sim N(0,\sigma_s^2 \Delta t + \sigma_r^2 \Delta t)$ [13], so based on the principle of linear combination of Gaussian variables, the distance variation during $\Delta t$ which is shown in (2) follows the Gaussian distribution as follows:

$$\Delta d(v_s, v_r)_{\Delta t} \sim N\left(0, \left(\sigma_s^2 + \sigma_r^2\right)\Delta t^3\right) \quad (9)$$

According to (7) and (9), since the $d_{sr,t_1}$ is constant during $\Delta t$, so the $d_{sr,\Delta t}$ shown in (7) also follows the Gaussian distribution, which is:

$$d_{sr,\Delta t} \sim N\left(d_{sr,t_1}, \left(\sigma_s^2 + \sigma_r^2\right)\Delta t^3\right) \quad (10)$$

Therefore, according to (7) and (8), the SINR of vehicle $v_r$ at $t_1 + \Delta t$ can be expressed as:

$$SINR_{r,t_1+\Delta t} = \frac{(d_{sr,t_1} + \Delta d(v_s, v_r)_{\Delta t})^{-\alpha}}{N + \sum_{i=1,i\neq s}^{n_{int-v}} (d_{ir,t_1} + \Delta d(v_s, v_r)_{\Delta t})^{-\alpha}} \quad (11)$$

Then the probability that after $\Delta t$, the SINR at the receiver vehicle $v_r$ is larger than the receiving threshold $\beta$ can be calculated as:

$$Pr_{sr}^{SINR}(\Delta t) = p\left\{ \frac{(d_{sr,t_1} + \Delta d(v_s, v_r)_{\Delta t})^{-\alpha}}{N + \sum_{i=1,i\neq s}^{n_{int-v}} (d_{ir,t_1} + \Delta d(v_s, v_r)_{\Delta t})^{-\alpha}} \geq \beta \right\} \quad (12)$$

As shown in (12), two parameters can affect the probability prediction of SINR: 1) the distances between the vehicle $v_r$ and its interference vehicles; 2) the number of the interference vehicles of $v_r$. With the time goes on, these two parameters change. On one hand, the interference vehicles of $v_r$ at $t_1$ may not the interference vehicles at $t_1 + \Delta t$, and the vehicles which are not the interference vehicles of $v_r$ at $t_1$ may be the interference vehicles at $t_1 + \Delta t$. As shown in Fig. 2, at $t_1$, the interference vehicles of $v_r$ are vehicles *A*, *B*, *C*, and *D*; at $t_1 + \Delta t$, the interference are vehicles *C*, *D*, *E*, *F*, and *G*. On the other hand, the distances between the vehicle $v_r$ and its interference vehicles change with the vehicle movement. Therefore, during the SINR prediction, both these two parameters need to be taken into account.

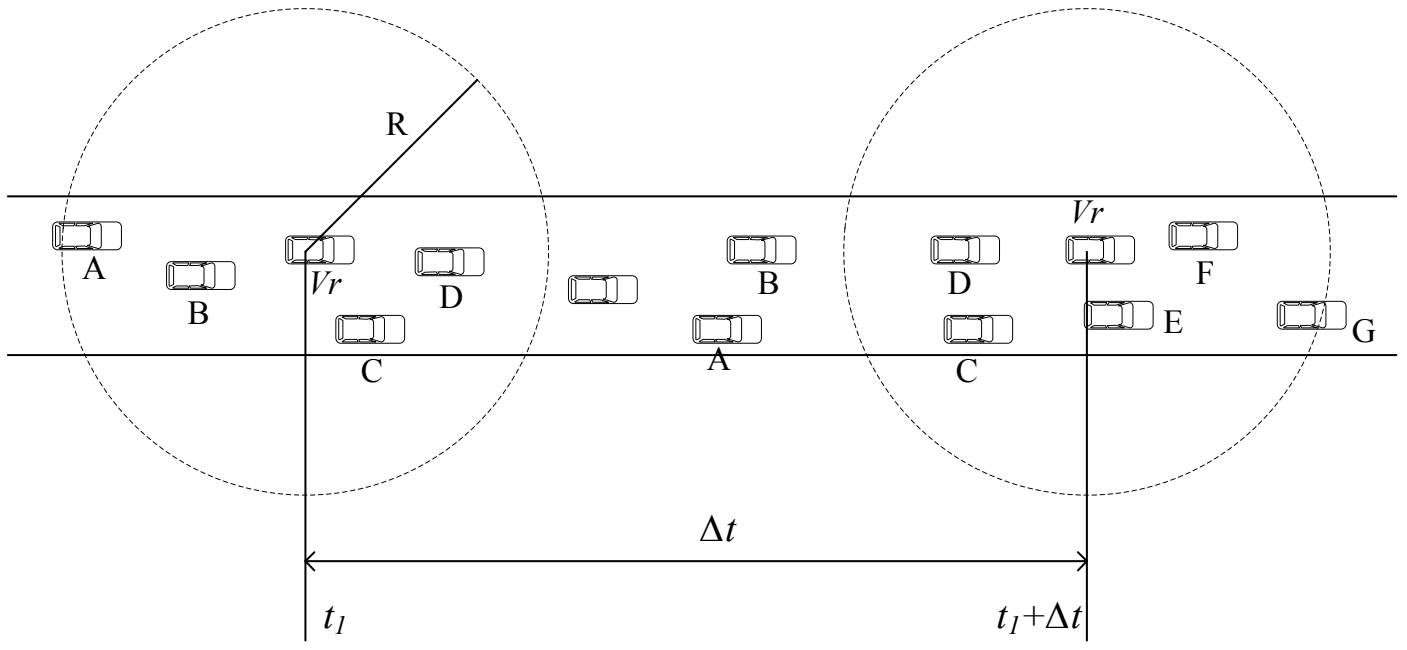

Fig. 2. Network topology variation

1. The effection of the distances between $v_r$ and its interference vehicles

For calculating the probability shown in (12), we need to know the probability density functions of $(d_{sr,t_1} + \Delta d(v_s, v_r)_{\Delta t})^{-\alpha}$ and $N + \sum_{i=1,i\neq s}^{n_{int-v}} (d_{ir} + \Delta d(v_s, v_r)_{\Delta t})^{-\alpha}$, respectively. Here we propose three theorems as follows to calculate the probability density functions of $(d_{sr,t_1} + \Delta d(v_s, v_r)_{\Delta t})^{-\alpha}$ and $N + \sum_{i=1,i\neq s}^{n_{int-v}} (d_{ir} + \Delta d(v_s, v_r)_{\Delta t})^{-\alpha}$.

**Theorem 1.** For $x \sim N(\mu, \sigma^2)$, the probability density function of $y = x^{-\alpha}$ can be expressed as:

$$f_Y(y) = y^{-\frac{1+\alpha}{\alpha}} \frac{1}{\sqrt{2\pi\alpha}} e^{-\frac{\left(y^{-\frac{1}{\alpha}} - \mu\right)^2}{2}}.$$

*Proof.* See Appendix A.

**Theorem 2.** For $x \sim N(\mu, \sigma^2)$, the probability density function of $z = N + \sum_{i=1}^{m} x_i^{-\alpha}$ can be expressed as:

$$f_z(z) = p(N) + \underbrace{\int_{-\infty}^{\infty} \cdots \int_{-\infty}^{\infty}}_{m} \left(z - \sum_{i=2}^{m} y_i\right)^{-\frac{1+\alpha}{\alpha}} \frac{1}{\sqrt{2\pi\alpha}} e^{-\frac{\left(\left(z - \sum_{i=2}^{m} y_i\right)^{-\frac{1}{\alpha}} - \mu\right)^2}{2}} \cdot \prod_{i=2}^{m} y_i^{-\frac{1+\alpha}{\alpha}} \frac{1}{\sqrt{2\pi\alpha}} e^{-\frac{\left(y_i^{-\frac{1}{\alpha}} - \mu\right)^2}{2}} dy_2 \cdots dy_m$$

*Proof.* See Appendix B.

**Theorem 3.** For $x \sim N(\mu, \sigma^2)$, $y = x^{-\alpha}$ and $z = N + \sum_{i=1}^{n} x_i^{-\alpha}$, then the probability density function of $w = y / z$ can be expressed as:

$$f_W(w) = \int_{-\infty}^{\infty} |z| (wz)^{-\frac{1+\alpha}{\alpha}} \frac{1}{\sqrt{2\pi\alpha}} e^{-\frac{\left((wz)^{-\frac{1}{\alpha}} - \mu\right)^2}{2}} \cdot f_Z(z) dz$$

where $f_z(z)$ can be calculated by Theorem 2.

*Proof.* See Appendix C.

As shown in Section IV.A, the probability density function of $d_{sr,\Delta t}$ can be calculated by (10); moreover, $d_{ir,\Delta t}$ follows the similar distribution as $d_{sr,\Delta t}$, which is $d_{ir,\Delta t} \sim N\left(d_{ir,t_1}, \left(\sigma_i^2 + \sigma_r^2\right)\Delta t^3\right)$. Therefore, according to Theorem

1, Theorem 2, and Theorem 3, the $Pr_{sr}^{SIR}(w)$ shown in (12) can be calculated as:

$$Pr_{sr}^{SINR}(w) = p\left\{\frac{(d_{sr}+\Delta d(v_s,v_r)_{\Delta t})^{-\alpha}}{N+\sum_{i=1,i\neq s}^{m}(d_{ir}+\Delta d(v_i,v_r)_{\Delta t})^{-\alpha}} \geq \beta\right\}$$
$$= 1 - p\left\{\frac{(d_{sr}+\Delta d(v_s,v_r)_{\Delta t})^{-\alpha}}{N+\sum_{i=1,i\neq s}^{m}(d_{ir}+\Delta d(v_i,v_r)_{\Delta t})^{-\alpha}} < \beta\right\} \quad (13)$$
$$= 1 - \int_0^{\beta}|z|(wz)^{-\frac{1+\alpha}{\alpha}}\frac{1}{\sqrt{2\pi}\alpha}e^{-\frac{\left((wz)^{-\frac{1}{\alpha}}-\mu\right)^2}{2}}\cdot f_Z(z)dz$$

where $f_z(z)$ can be calculated by Theorem 2.

However, the probability shown in (13) does not take the number variation of the interference vehicles into account. Due to the high mobility of vehicles, the number of interference vehicles of $v_r$ changes greatly (as shown in Fig. 2), which can affect the SINR seriously. In the following, we will calculate the effection of the number of interference vehicles on the probability prediction of SINR.

2. The effection of the number of interference vehicles

The number of interference vehicles relates to the distances between the receiver $v_r$ and its interference vehicles, which can affect the SINR of the receiver greatly. The interference vehicles of $v_r$ at $t_1$ may not the interference vehicles at $t_1+\Delta t$, and the vehicles which are not the interference vehicles of $v_r$ at $t_1$ may be the interference vehicles at $t_1+\Delta t$, which has been illustrated in Fig. 2.

After $\Delta t$, the probability that the vehicle $v_r$ locates in the transmission range of vehicle $v_s$ is $P_{sr}^{d}(\Delta t)$, which can be calculated by (3). Therefore, assuming that there are $n$ vehicles in the network, then at $t_1+\Delta t$, the average number of neighbors of $v_r$ can be calculated as:

$$n_{\Delta t} = \sum_{i=1}^{n} P_{ir}^{d}(\Delta t) \quad (14)$$

According to Theorem 2 and (14), after $\Delta t$, the probability distribution function of the SINR at the receiver $v_r$ can be calculated as:

$$f_z(z) = p(N) + \underbrace{\int_{-\infty}^{\infty}\cdots\int_{-\infty}^{\infty}}_{n_{\Delta t}}\left(z-\sum_{i=2}^{n_{\Delta t}}x_i\right)^{-\frac{1+\alpha}{\alpha}}\frac{1}{\sqrt{2\pi}\alpha}e^{-\frac{\left(\left(z-\sum_{i=2}^{n_{\Delta t}}x_i\right)^{-\frac{1}{\alpha}}-\mu\right)^2}{2}}$$
$$\cdot\prod_{i=2}^{n_{\Delta t}}x_i^{-\frac{1+\alpha}{\alpha}}\frac{1}{\sqrt{2\pi}\alpha}e^{-\frac{\left(x_i^{-\frac{1}{\alpha}}-\mu\right)^2}{2}}dx_2\cdots dx_{n_{\Delta t}} \quad (15)$$

Then the probability $Pr_{sr}^{SIR}(w)$ can be calculated by (13) and (15). Note that in this calculation, the value of $f_z(z)$ has changed. In (15), both the distances between the vehicle $v_r$ and its interference vehicles and the number of the interference vehicles of $v_r$ are taken into account.

*B. Prediction algorithm of PQL*

In this section, the probability prediction algorithm of PQL has been proposed. From the viewpoint of PQL, if the sender vehicle $v_s$ can send data packet to the receiver vehicle $v_r$ successfully, the receiver vehicle $v_r$ should have enough memory space to store the data packet that transmitted from the sender vehicle $v_s$, which means that the PQL at the receiver $v_r$ mush smaller than the maximum allowed value.

Assuming that there are $n_{t_1}$ neighbor vehicles of receiver $v_r$ at $t_1$, then after $\Delta t$, this number can be calculated by (14). However, the neighbors at both $t_1$ and $t_1+\Delta t$ can send data packet to the receiver vehicle, so there are two kinds of neighbor vehicles: 1) the vehicles which locate in the transmission range of receiver $v_r$ at $t_1$, such as vehicles *A*, *B*, *C*, and *D* in Fig. 2; 2) the vehicles which do not locate in the transmission area of $v_r$ but move into it during $\Delta t$, such as the vehicles *E*, *F*, and *G* in Fig. 2. The PQL prediction should take these two kinds of neighbor vehicles into account.

Since the probability that a vehicle has data packet need to be transmitted relates to the MAC protocols and different MAC protocols have different generation probabilities (which is not the main research topic of this paper), therefore, without loss of generate, we use $p_0$ to represent the data packet generation probability in this paper. The detail of how the MAC protocols determine the generation probability of vehicles can be found in [44] and [45].

Based on these assumptions, we assume that the data generation probability of vehicle is $p_0$; the time interval is $\Delta t$; the transmission interval in the MAC protocol is $t_m$; therefore, the number of transmission intervals is $n_{inter} = \Delta t / t_m$. For calculating the probability that after $\Delta t$, the PQL is smaller than the maximum allowed value, we assume that the maximum allowed PQL is *M* and the PQL is *a* at $t_1$. So the remaining available PQL is $b = M - a$. In each transmission interval, only one vehicle can transmit one data packet; the transmitter could be the receiver vehicle $v_r$ or its neighbor vehicles. When the sender is the neighbor vehicle of $v_r$, the PQL in $v_r$ increases; otherwise, if the sender is $v_r$, then the PQL decreases. If the receiver $v_r$ can receive data packet transmitted by sender $v_s$ after $\Delta t$, the PQL in $v_r$ should smaller than the maximum PQL and the number of data packets received by $v_r$ during $\Delta t$ should less than the number of data packets sent by $v_r$. In each transmission interval, there are three different situations: 1) the neighbor vehicles of $v_r$ send data packet; 2) the $v_r$ sends the data packet; 3) both the neighbor vehicles and $v_r$ do not send data packet, which can be found in Fig. 3.

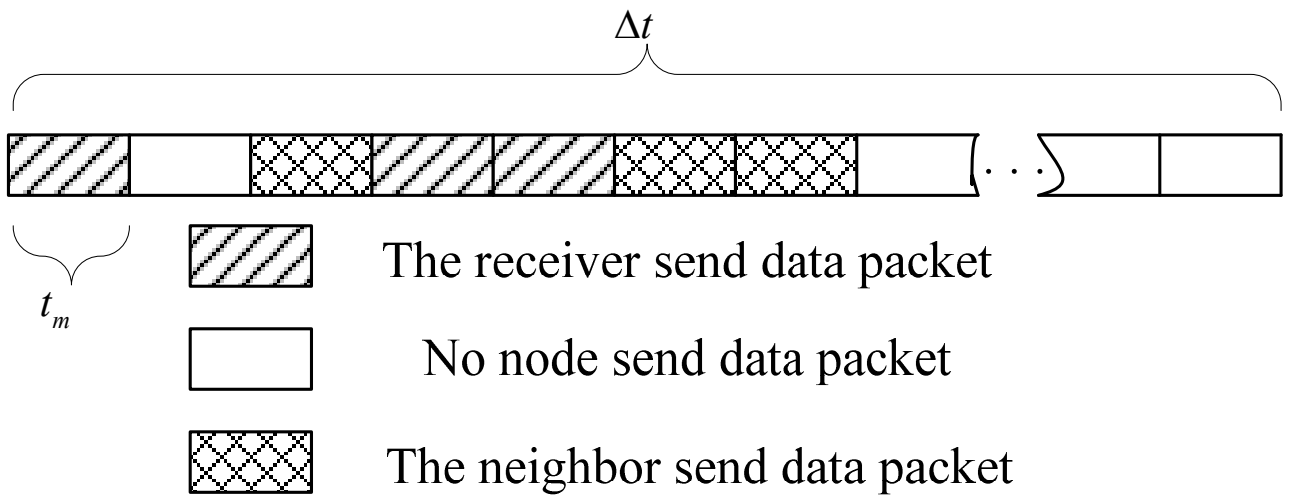

Fig. 3. Different transmission situations during the transmission interval

For the first situation, assuming that there are $n_{\Delta t_j}$ neighbor vehicles of $v_r$ in the *jth* transmission interval, then the probability that there is at least one neighbor vehicle generates data packet can be calculated as:

$$p_{nei} = 1-(1-p_0)^{n_{\Delta t_j}} \tag{16}$$

Similarly, the probability of situation 2 and situation 3 can be calculated as:

$$p_{rec} = p_0 \tag{17}$$

$$p_{none} = (1-p_0)^{n_{\Delta t_j}+1} \tag{18}$$

The number of neighbors at *jth* transmission interval can be calculated based on (14), which is:

$$n_{\Delta t_j} = \sum_{i=1}^{n} P_{ir}^{d}(\Delta t_j),\ j=1,2,...,n_{inter} \tag{19}$$

where $\Delta t_j = t_1 + jt_m$. So the average number of neighbors during $\Delta t$ can be calculated as:

$$\overline{n}_{\Delta t} = \frac{1}{n_{inter}} \sum_{j=1}^{n_{inter}} \sum_{i=1}^{n} P_{ir}^{d}(\Delta t_j) \tag{20}$$

As shown in Fig. 3, since during some transmission intervals, there may no vehicles send data packet, so the average number of transmission intervals in which there have a data packet need to be transmitted can be calculated as:

$$\overline{n}_{inter} = n_{inter} - \sum_{j=1}^{n_{inter}} (1-p_0)^{\overline{n}_{\Delta t}+1} \tag{21}$$

where $\sum_{j=1}^{n_{inter}} (1-p_0)^{\overline{n}_{\Delta t}+1}$ is the average number of transmission intervals in which there have no data packet needed to be transmitted. According to (20), during $\Delta t$, the number of received data packets $x$ and the sent data packets $y$ by $v_r$ should meet the requirements as follows:

$$\begin{cases} x+y=\overline{n}_{inter} \\ x-y \le b \end{cases} \tag{22}$$

where $b$ is the remaining available PQL. According to (22), we can conclude that the maximum and minimum number of data packet that can be transmitted by the neighbor vehicle of $v_r$ are $x=(\overline{n}_{inter}+b)/2$ and $x=(\overline{n}_{inter}-b)/2$, respectively. Therefore, the probability that after $\Delta t$, the packet queue length of $v_r$ is smaller than the maximum allowed packet queue length can be calculated as:

$$Pr_r^Q(\Delta t) = \sum_{i=0}^{(\overline{n}_{inter}+b)/2} C_{\overline{n}_{inter}}^{i} \left(1-(1-p_0)^{\overline{n}_{\Delta t}}\right)^{i} \cdot (1-p_0)^{\overline{n}_{inter}-i} \tag{23}$$

In (23), $\overline{n}_{inter} = n_{inter} - \sum_{j=1}^{n_{inter}} (1-p_0)^{\overline{n}_{\Delta t}+1}$, $n_{inter} = \Delta t / t_m$, and $\overline{n}_{\Delta t} = \frac{1}{n_{inter}} \sum_{j=1}^{n_{inter}} \sum_{i=1}^{n} P_{ir}^{d}(\Delta t_j)$.

Based on Section IV.A and Section IV.B, after $\Delta t$, the probability that the SINR of the receiver is larger than the receiving threshold and the probability that the PQL of the receiver is smaller than the maximum allowed value can be calculated. The next step is to select the relaying vehicles in the candidate set for each sender based on these predicted probabilities.

## V. PROBABILITY PREDICTION BASED HIGH-EFFICIENT AND BALANCED OPPORTUNISTIC ROUTING

In this section, we will propose the probability prediction based reliable opportunistic routing (PRO) algorithm based on the conclusions in Section IV. Since the PRO algorithm is geographic based algorithm, so in this paper, only the neighbor vehicles which distances to the destination vehicle are smaller than that of the sender and move toward to the destination can be chosen as the candidate relaying vehicles. The set of the candidate relaying vehicles is defined as the candidate set.

### *A. Vehicle utility calculation algorithm*

When the sender gets $Pr_{sr}^{SIR}(w)$ and $Pr_r^Q(\Delta t)$ of the candidate relaying vehicles, then the utilities of these relaying vehicles needed to be determined based on these two parameters. In the candidate set, each relaying vehicle can be expressed by $Pr_{sr}^{SIR}(w)$ and $Pr_r^Q(\Delta t)$, i.e. $node_i = \left\{Pr_{sr(i)}^{SINR}(w), Pr_{r(i)}^{Q}(\Delta t)\right\}$. Assuming that there are $n$ candidate relaying vehicles, so the set of $Pr_{sr}^{SIR}(w)$ and $Pr_r^Q(\Delta t)$ of different relaying vehicles are $P_{SINR}(n) = \left\{Pr_{sr(1)}^{SINR}(w), Pr_{sr(2)}^{SINR}(w), ..., Pr_{sr(n)}^{SINR}(w)\right\}$ and $P_Q(n) = \left\{Pr_{r(1)}^{Q}(\Delta t), Pr_{r(2)}^{Q}(\Delta t), ..., Pr_{r(n)}^{Q}(\Delta t)\right\}$, respectively.

When calculating the utilities of the relaying vehicles, the ideal situation is that the relaying vehicle which the utility is the highest has highest $Pr_{sr}^{SIR}(w)$ and $Pr_r^Q(\Delta t)$; the vehicle which the utility is the second highest has the second highest $Pr_{sr}^{SIR}(w)$ and $Pr_r^Q(\Delta t)$, and so on. However, this is not always feasible. The most common situation is that the relaying vehicle has excellent performance at one aspect and ordinary performance on the other aspect, which can be found in Fig. 4. For instance, in Fig. 4, the first parameter in node 2 is high while the second parameter is low.

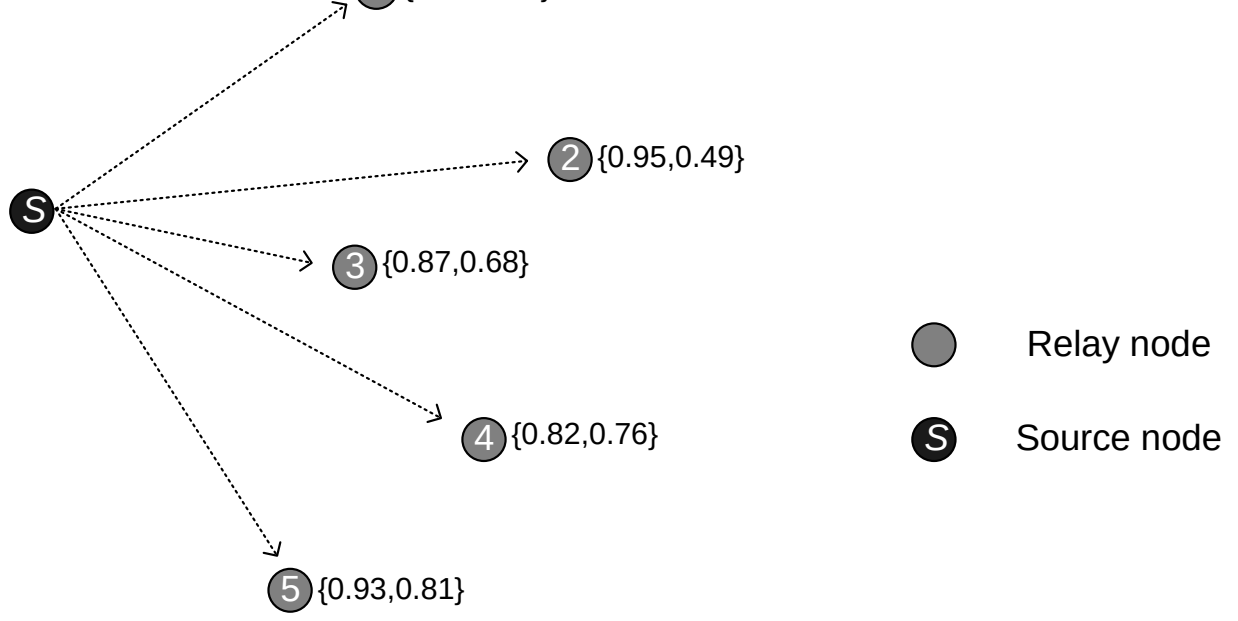

Fig. 4. The parameters of the relaying vehicles in opportunistic routing

Therefore, during calculating the utilities of the relaying vehicles, both of these two parameters should be taken into account. There is a fact that for the parameter $Pr_{sr}^{SIR}(w)$ and $Pr_r^Q(\Delta t)$, the effection of these two parameters on the routing performance is not same. The parameter which the variance is larger will have greater effection on the routing performance than that of the parameter which the variance is smaller. For the parameter which the variance is large, to different relaying vehicles, the routing performance changes greatly; for the parameter which the variance is small, this changing is slight. For instance, in Table 1, the variance of $Pr_{sr}^{SIR}(w)$ is much larger than that of $Pr_r^Q(\Delta t)$; therefore, during the relaying vehicle selection, for $Pr_r^Q(\Delta t)$, which vehicle is chosen has small effection on the routing performance, since the difference between these four nodes are quite small; however, to $Pr_{sr}^{SIR}(w)$, which vehicle is chosen as the first relaying node will have great effection on the routing performance. For example, for node 3 and node 4, even the $Pr_r^Q(\Delta t)$ of node 4 is larger than that of node 3, the utility of node 3 should larger than that of node 4, since the $Pr_{sr}^{SIR}(w)$ (which the variance is much larger than that of $Pr_r^Q(\Delta t)$) of node 3 is much larger than that of node 4.

Table 1. Parameter with different variance

| | node 1 | node 2 | node 3 | node 4 | variance |
|---|---|---|---|---|---|
| $Pr_{sr}^{SIR}(w)$ | 0.11 | 0.34 | 0.67 | 0.49 | 0.056 |
| $Pr_r^Q(\Delta t)$ | 0.81 | 0.83 | 0.815 | 0.824 | $8.2\times10^{-5}$ |

Based on the analysis above, in this paper, we introduce the weight based approach into the calculation of the utility of vehicle. The weight represents the effection of the parameter on the routing performance. Since the parameter which the variance is larger has greater effection on the routing performance than that of the smaller one, so in this paper, we use the variances of $P_{SINR}(n)$ and $P_Q(n)$ as the weights to calculate the vehicle utility, which can be calculated as:

$$U = v_{SINR}\cdot Pr_{sr}^{SINR}(w) + v_Q\cdot Pr_r^Q(\Delta t) \tag{24}$$

where $v_{SINR}$ is the variance of $P_{SINR}(n)$, and $v_Q$ is the variance of $P_Q(n)$. For evaluating the difference between the variances of these two parameters, we define the parameter resolution ratio $\xi$ as:

$$\xi = \begin{cases} \dfrac{v_{SINR}}{v_Q}, & v_{SINR} > v_Q \\ 1, & v_{SINR} = v_Q \\ \dfrac{v_Q}{v_{SINR}}, & v_Q > v_{SINR} \end{cases} \tag{25}$$

From (25), we can find that $\xi \geq 1$, the larger $\xi$ is, the larger difference between the variances of these two parameters.

For the vehicle utility calculated in (24), with the increasing of $\xi$, the effection of the parameter which the variance is large on the vehicle utility increases, and the effection of the parameter which the variance is small decreases. When the $\xi$ is small, the effection of these two parameters on the vehicle utility is similar. For instance, as the parameters shown in Fig. 5, the parameter resolution ratio is $\xi = 6.61$, which is much larger than 1. So the utility of vehicle 1 (in which the $Pr_r^Q(\Delta t)$ is the largest) is larger than that of vehicle 3 and vehicle 2. However, as shown in Fig. 6, in which the $\xi = 1.1$, the results are different. In Fig. 6(a), the utilities of the relaying vehicles are presented; the relaying vehicle which the utility is the largest also has the largest $Pr_{sr}^{SIR}(w)$. So it seems that the vehicle utility is decided by $Pr_{sr}^{SIR}(w)$ which the variance is larger. However, as shown in Fig. 6(b), if the $Pr_r^Q(\Delta t)$ is exchanged between different vehicles, the priority of the same vehicle changes. The first priority relaying node in Fig. 6(b) becomes the second priority relaying node in Fig. 6(a), in which the $Pr_{sr}^{SIR}(w)$ is not the largest. The results shown in Fig. 6 demonstrate that when the $\xi$ is small, the vehicle utility will be determined by both of these two parameters. However, to the parameters shown in Fig. 5, if the parameters are exchanged, the priorities of the vehicles are also determined by $Pr_r^Q(\Delta t)$.

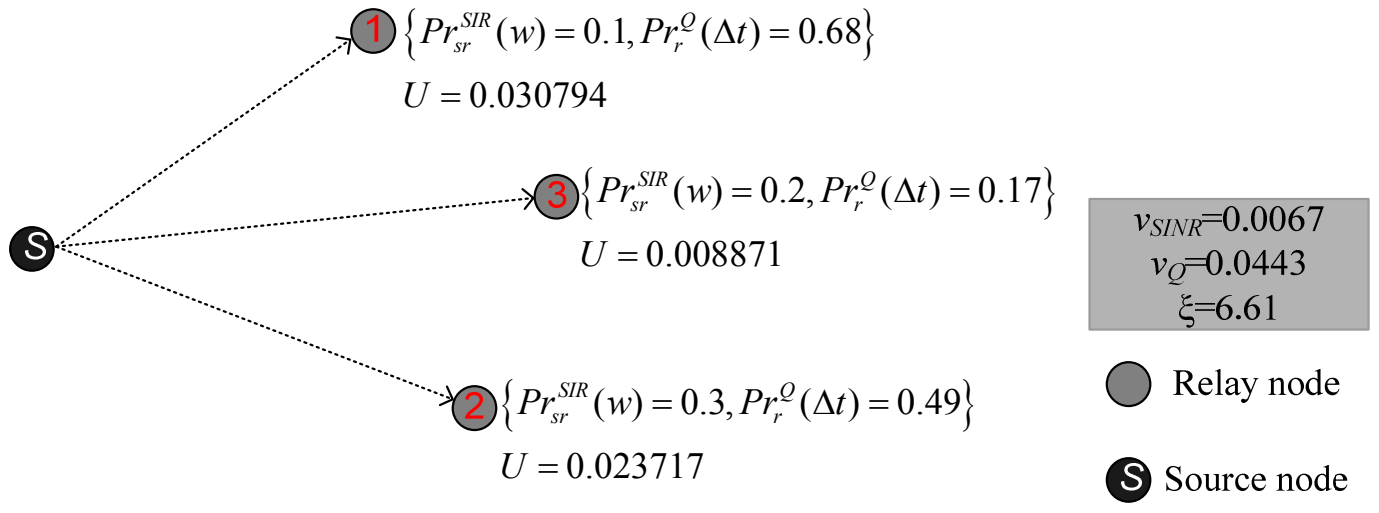

Fig. 5. Vehicle utility and relaying priority when the $\xi$ is large

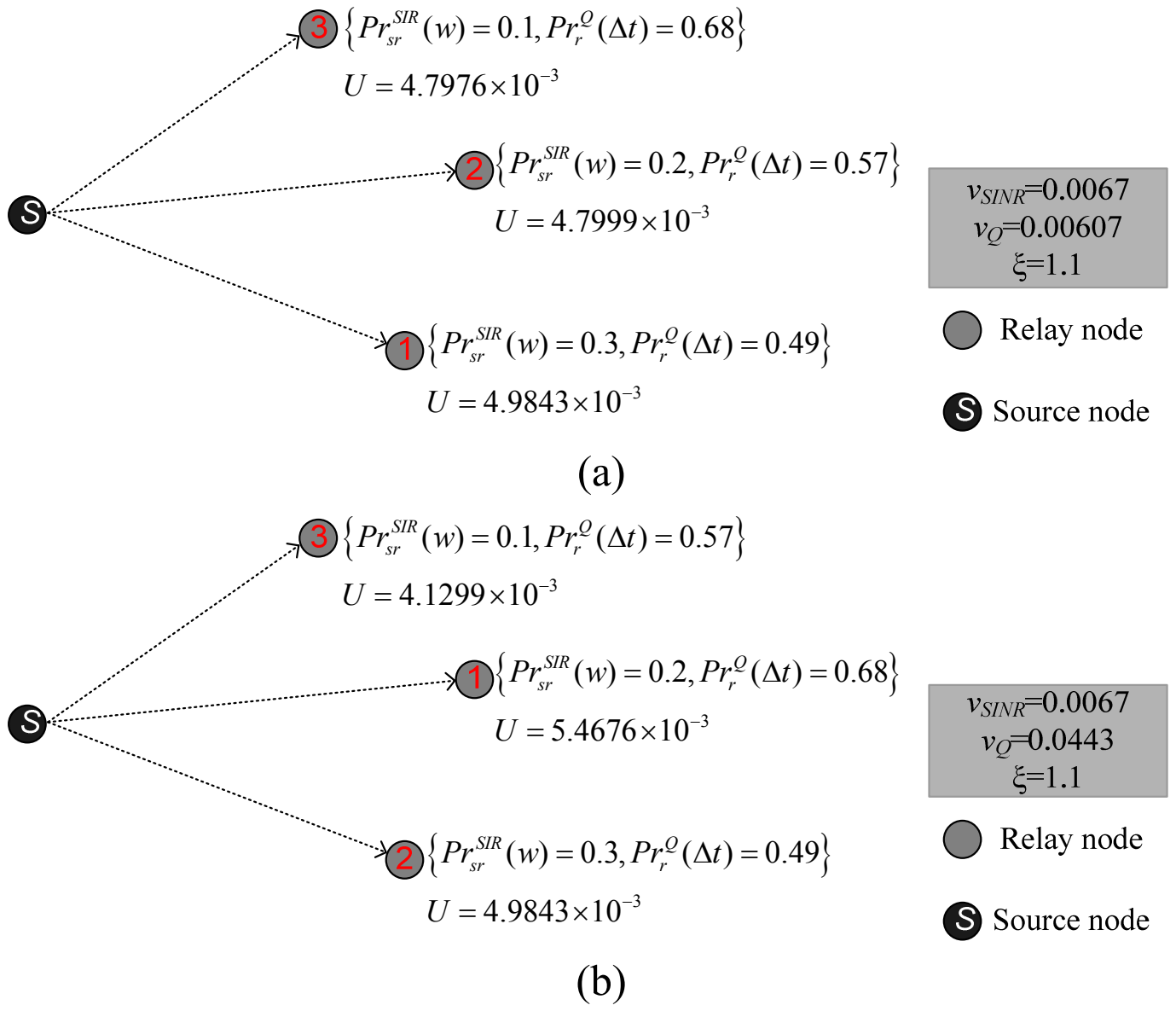

Fig. 6. Vehicle utility and relaying priority when the $\xi$ is small; the $Pr_r^Q(\Delta t)$ has been exchanged between the first and second vehicle in (a) and (b).

### *B. Candidate relaying node set optimization*

In opportunistic routing, the number of vehicles in the candidate set is important to the routing performance. On one hand, the more vehicles in the candidate set, the higher packet delivery ratio is; however, when there are too many vehicles in the candidate set, the duplicate transmission and the interference increase. Therefore, the number of vehicles in the candidate set should be optimized and the inappropriate vehicles should be removed from the candidate set. The packet delivery ratio between the sender and the candidate set is defined as the probability that the data packet sent by the sender can be received successfully by at least one relaying vehicles in the candidate set, which can be calculated as:

$$P_{opp} = 1 - \prod_{i=1}^{n_{rel}} (1 - P_i) \tag{26}$$

where $n_{rel}$ is the number of vehicles in the candidate set; $P_i$ is packet delivery ratio of *ith* vehicle in the candidate set. Assuming that the threshold of the packet delivery ratio is $P_{opp}^*$, where $P_{opp}^*$ is an application-specify parameter and decided by different application requirements, then based on (26), the number of vehicles $n_{rel}$ in the candidate set should satisfy the constraint that $P_{opp} \geq P_{opp}^*$. However, as shown in [46] and [47], for minimizing the interference and duplicate transmission, the number of vehicles in the candidate set should be limited; therefore, the minimum $n_{rel}$ got from (26) will be chosen as the number of vehicles in the candidate set, noted as $n_{rel}^*$. When the number of relaying vehicles has been calculated, then the first $n_{rel}^*$ vehicles in the candidate set will be chosen as the relaying vehicles. If the number of vehicles in the candidate set is larger than $n_{rel}^*$, then the redundant relaying vehicles (i.e. the last $n_{rel} - n_{rel}^*$ vehicles in the candidate set) will be removed from the candidate set; if the number of vehicles in the candidate set is smaller than $n_{rel}$, then all the vehicles in the candidate set will be chosen as the relaying vehicles. Moreover, the relaying priority of each vehicle will be assigned based on the vehicle utility. The rule for determining the relaying priority is: the larger utility, the higher relaying priority is. Some examples can be found in Fig. 5 and Fig. 6, the vehicle utilities and the relaying priorities are presented in these figures.

### C. Routing process

When the sender vehicle want send data packet, first, it selects the candidate set based on the geographic information of its neighbors; only the neighbors which the distances to the destination vehicle are smaller than that of the sender vehicle can be selected as the candidate relaying vehicles. The sender vehicle predicts the $Pr_{sr}^{SIR}(w)$ and $Pr_r^Q(\Delta t)$ for each vehicle in the candidate set and calculates the utilities of the candidate relaying vehicles based on the weight based algorithm. Based on the candidate set optimization algorithm that introduced in Section IV.B, the candidate set is optimized. Then the sender vehicle broadcasts the data packet to all the relaying vehicles in the candidate set which has been optimized. This data packet includes the candidate set and the relaying priority of each relaying vehicle in this set.

When the relaying vehicles receive the data packet transmitted from the sender vehicle, then the relaying priority based relaying algorithm introduced in [48] will be applied. In the relaying priority based relaying algorithm, each neighbor vehicle monitors the packet transmitted from the sender vehicle. When the neighbor vehicle receives the data packet, first it checks if it is included in the candidate set. If not, it discards the packet directly. Otherwise, it sets its forwarding timer as follows. The *ith* relaying vehicle on the candidate set sets its forwarding timer to $(i-1)T$, where $i$ is the relaying priority of the relaying vehicles in the candidate set and starts from 1. In this way, the vehicle with larger utility forwards the packet earlier, and other nodes hearing its forwarding will cancel their forwarding timer and remove the packet from their packet queue, thereby avoiding duplicate forwarding. In [48], the waiting time $T$ is 45*ms*, which is appropriate for bulk transfer, targeted by all opportunistic routing; so in this paper, we use the same waiting time as [48].

The process of the PRO routing algorithm can be expressed below.

**Algorithm 1:** Probability prediction based reliable opportunistic routing (PRO) algorithm

Notations:
*s*: The sender vehicle;
*Packet(i)*: The *ith* packet in the sender *s*;
$d_{node(s)}$: the distance of sender vehicle *s* to the destination node;
$d_{node(i)}$: the distance of the *ith* neighbor of sender *s* to the destination node;
$\mathbb{R}_s$ : the candidate relaying set;
$\mathbb{R}_s^*$ : the optimized relaying set;
*node(i)*: the *ith* relaying nodes in $\mathbb{R}_s$ ;
*move(i)*: the moving direction of *ith* candidate relaying node;
*direction(d)*: the direction of the destination node *d*;
$U_i$: the node utility of *ith* relaying nodes in $\mathbb{R}_s$ ;
$T_{node(i)}$: the timer of relaying node *i*;
$T_wait_{node(i)}$: the waiting time of relaying node *i* before receiving the ACK from the higher priority relaying node;

1. **for** *Packet(i)* **do**
2. **if** $d_{node(i)} \leq d_{node(s)}$ && *move(i)* == *direction(d)* **then**
3. $\mathbb{R}_s \leftarrow$ *node(i)*;
4. **end if**
5. Predicating the $Pr_{sr}^{SIR}(w)$ and $Pr_r^Q(\Delta t)$ for each node in $\mathbb{R}_s$ ;
6. Calculating the $v_{SINR}$ and $v_Q$ of $P_{SINR}(n)$ and $P_Q(n)$;
7. Calculating the node utility $U_i$ for each node in $\mathbb{R}_s$ based on (24);
8. Assigning the priorities to the nodes in $\mathbb{R}_s$ based on $U_i$;
9. Calculating the optimized number of node $n_{rel}$ in $\mathbb{R}_s$ ;
10. **if** $n > n_{rel}$ **then**
11. remove the last $n_{rel}$ − $n$ relaying nodes from $\mathbb{R}_s$ ;
12. **else**
13. keep all the relaying nodes in $\mathbb{R}_s$ ;
14. **end if**
15. Updating the relaying node set as $\mathbb{R}_s^*$ ;
16. The sender node broadcasts the data packet with the relaying priority list *L(s)* to the nodes in $\mathbb{R}_s^*$ ;
17. **if** *node(i)* $\in$ *L(s)* **then**
18. *node(i)* receive the data packet;
19. $T_{node(i)} = (k-1)T$;
20. **else**
21. *node(i)* drop the data packet;
22. **end if**
23. **if** $T_wait_{node(i)} = T_{node(i)}$ **then**
24. node *i* relaying the data packet to the next hop relaying nodes the same as the Step 1 to Step 16;
25. **else if** $T_wait_{node(i)} < T_{node(i)}$ **then**
26. node *i* drop the data packet;
27. **end if**
28. **end for**.

## VI. SIMULATION AND DISCUSSION

### A. Network configuration

In this section, for evaluating the performance of the probability prediction reliable opportunistic routing algorithm, we compare the PRO algorithm with GPSR, ExOR, and SRPE. Since the performance comparison between the GPSR, ExOR, and SRPE has been done by [13] and [19], therefore, for getting more fair results, the network configuration in this paper is similar to that shown in [13] and [19]. The simulation parameters can be found in Table 2.

Table 2

| simulation parameter | value |
|---|---|
| simulation area | 2000*m*×2000*m* |
| number of vehicles | 100, 150,…, 300 |
| transmission range | 250*m* |
| channel data rate | 2Mbps |
| the traffic type | Constant Bit Rate (CBR) |
| number of CBR connection pairs | 20, 40,…, 100 |
| packet size | 512bytes |
| minimum velocity | 30*km/h* |
| maximum velocity | 60*km/h* |
| beacon interval | 1*s* |
| maximum packet queue length | 50 packets |
| propagation model | *Nakagami-m* model[42] |
| MAC layer | IEEE 802.ll DCF |
| simulation tool | NS2 |

The varying parameters during the simulation are the number of vehicles in the network and the number of CBR connection pairs [13][14][19]. During the simulation, the performance matrixes used in this paper are: (1) Packet delivery ratio: the packet delivery ratio is defined as the ratio of the number of packets received successfully by the destination vehicle to the number of packets generated by the source vehicle [13][19]; (2) End-to-End delay: the transmission delay of the data packet from the source vehicle to the destination vehicle; (3) Network throughput: the network throughput is the ratio of the total number of packets received successfully by the destination vehicle to the number of packets sent by all the vehicles during the simulation time [49]. The routing algorithms evaluated in this section are: GPSR routing algorithm [31], ExOR (street-centric) routing algorithm (street-centric ExOR can be explained as the opportunistic routing in which the basic unit is the sub-network rather than the single vehicle) [14][19], SRPE routing algorithm [13], and PRO algorithm.

*6.2 Performance under different node density*

In this section, the effection of different network densities on the routing performance will be evaluated. The number of CBR connection in this section is set to 20 and the data generation rate is 1 packet per second. The results of the routing performance can be found in Fig. 7, Fig. 8, and Fig. 9.

Fig. 7 illustrates the performance of the packet delivery ratio of these four routing algorithms under different node densities. With the increasing of the node density, the packet delivery ratio increases both in these four algorithms. This is due to the fewer vehicles in the network, the higher probability of network partition is, which means the communication links between different vehicles are easy to break; therefore, the packet delivery ratio is low when the network is sparse. When the network density increases, at the beginning, the packet delivery ratio increases fast. However, when the network density is large enough, the increasing becomes slow. For instance, when the vehicle number is 150 in the network, the packet delivery ratio increases greatly compared with that when the vehicle number is 100; however, when the vehicle number is 300, this increasing is small compared with that when the vehicle number is 250. The reasons can be explained as: 1) when the network is sparse, the probability of network partition is high, this probability reduces with the increasing of the node density; however, when the node density is large enough, the network will fully connect, then even the node density increases, the effection of vehicle number on the packet delivery ratio is slight; 2) with the increasing of the node density, the network interference and competition become more serious than that in the sparse network, which also limits the further routing performance improvement. Moreover, in Fig. 7, the packet delivery ratio variation of GPSR, ExOR, and SRPE are larger than that of PRO algorithm. This is easy to understand. Because the SINR and packet queue length are taken into account in PRO algorithm when determining the candidate set, so the packet delivery ratio in PRO algorithm is better than the other three algorithms; for instance, the packet delivery ratio in PRO algorithm is about 20% higher than SRPE when the number of vehicle is 100.

Not only the packet delivery ratio, but also the performance of end-to-end delay in PRO algorithm is better than the other three algorithms, which can be found in Fig. 8. In Fig. 8, when the number of vehicle is 100, the end-to-end delay of PRO algorithm is 3s; however, this value is 7s in SRPE algorithm and is 9s in GPSR algorithm, respectively. This can be explained by the performance of packet delivery ratio which is shown in Fig. 7. The high packet delivery ratio means low probability of retransmission and packet loss, which also contributes to reduce the end-to-end transmission delay. When the number of vehicles in the network increases, more transmission links can be chosen when send data packet to the same vehicle. So with the increasing of the node density, the end-to-end delay in these four algorithms reduces. Moreover, similar to the packet delivery ratio, when the network density is small, this reducing is obviously; with the increasing of the vehicle number, this reducing becomes slow. For instance, when the vehicle number increases from 100 to 150, the decreasing of the end-to-end delay is 3s in SRPE algorithm and is 2s in PRO algorithm, respectively; however, this decreasing is near to 0 in SRPE algorithm and PRO algorithm when the vehicle number increases from 250 to 300. This is because when the node density is small, the network partition is serious; so once the number of vehicles increases, the end-to-end delay will be reduced greatly. However, when the node density is large enough, the network partition will not be the determinant parameter of the end-to-end delay, the network interference and competition will affect the routing performance seriously; therefore, even the network density increases, the performance of end-to-end delay is not improved prominent.

The excellent performance of packet delivery ratio and end-to-end delay in PRO algorithm also contributes to the performance improvement of network throughput, which can be found in Fig. 9. In Fig. 9, the network throughput of the PRO algorithm is much better than that of the other three algorithms. For instance, the network throughput of PRO algorithm is about 18% higher than that of SRPE algorithm. Moreover, the network throughputs of these four algorithms are all stable. The reasons are: 1) when the node density increases, the probability of network partition decreases, so the network throughput increases; 2) when the node density increases, on one hand, the transmission hops to the destination vehicles increase, which reduces the network throughput; on the other hand, when the node density increases, the network interference and competition increase, which also reduces the performance of network throughput. When the number of vehicle is small, the first reason is the leading role; when the number of vehicle is large, the second reason has the main effection on the routing performance; so the network throughput is stable.

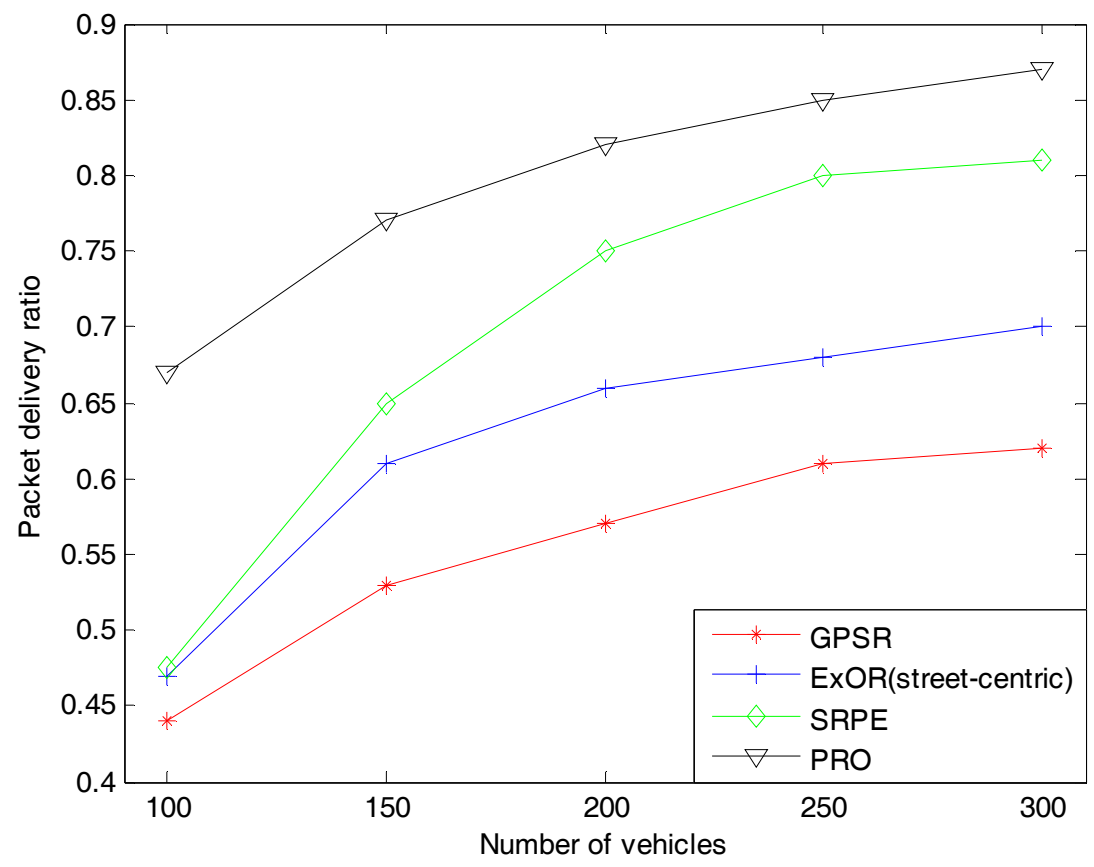

Fig. 7. Packet delivery ratio under different number of vehicles

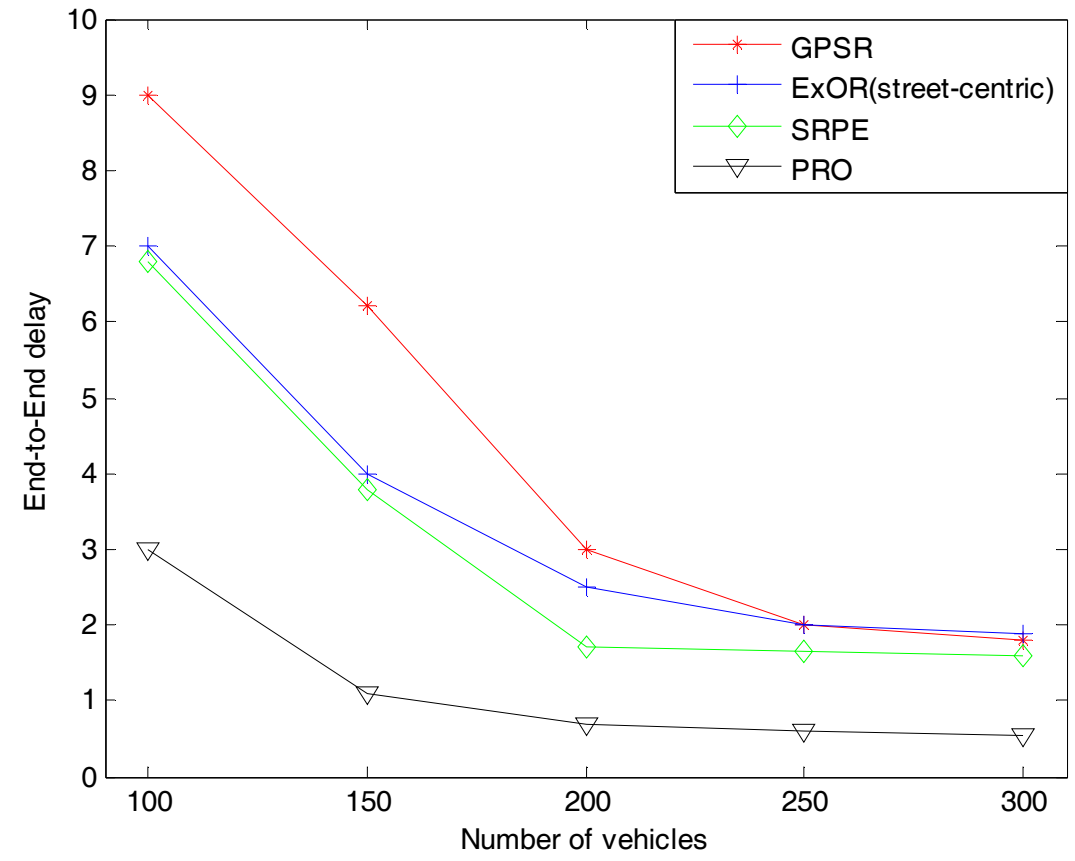

Fig. 8. End-to-End delay under different number of vehicles

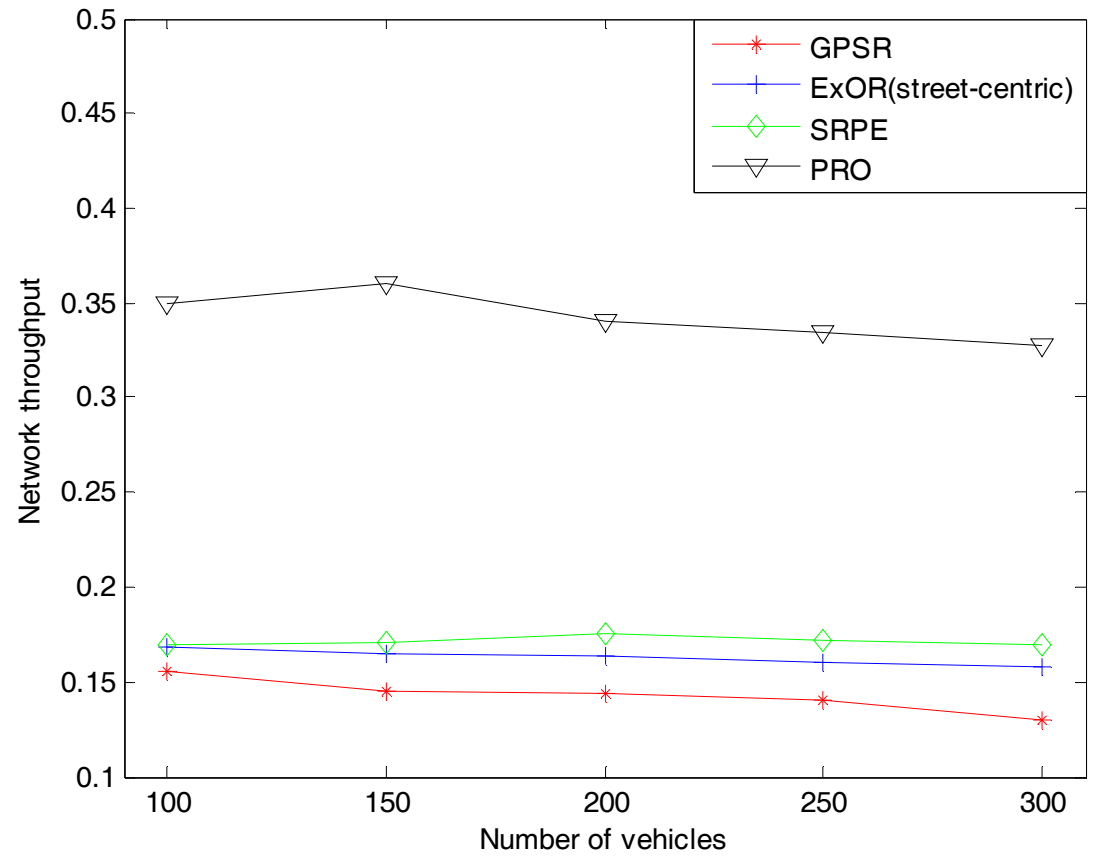

Fig. 9. Network throughput under different number of vehicles

### *6.3 Performance under different traffic load*

In this section, we will evaluate the performance of these four routing algorithms under different number of CBR connection pairs. The number of vehicles in the network is 200 and the data generation rate is 1 packet per second. As shown in [13], [17], and [19], the different number of CBR connection pairs represent different traffic load. The results have been shown in Fig. 10, Fig. 11, and Fig. 12.

In Fig. 10, the packet delivery ratios of these four routing algorithms under different number of CBR connection pairs have been demonstrated. With the increasing of the number of CBR connection pairs, the packet delivery ratios of these four algorithms decrease. When the number of CBR connection pairs is smaller than 60, the decreasing is slow (this number is 40 in GPSR algorithm); however, the decreasing is stable in PRO algorithm. This is due to two reasons: 1) when the traffic load increases, the network interference and competition increase, which causes the increasing of the packet loss and retransmission; so the packet delivery ratio deteriorates; 2) with the increasing of the traffic load, the probability of buffer overflow increases, which causes high probability of packet loss. However, due to the PRO routing algorithm takes the SINR and PQL into account during the routing decision, so on one hand, the performance of packet delivery ratio is much better than the other three routing algorithms; for instance, the packet delivery ratio in PRO algorithm is about 13% higher than that in SRPE algorithm when the number of CBR connection pairs is 100; on the other hand, the decreasing of the packet delivery ratio is stable in PRO algorithm; this is different with the other three routing algorithms which have sharply inflection points (60 in ExOR and SRPE, 40 in GPSR).

The SINR and the packet queue length can not only affect the packet delivery ratio, but also the end-to-end delay. As shown in Fig. 11, with the increasing of the traffic load, the end-to-end delay decreases when the number of CBR connection pairs is smaller than 60, and increase when this number is larger than 60. The effection of the traffic load on end-to-end delay is different with that of the packet delivery ratio (which is shown in Fig. 8). This can be explained as follows: when the traffic load is light, on one hand, the network interference and the probability of buffer overflow are small; on the other hand, even these parameters increase with the increasing of the traffic load, the network capability is far away from the saturation state; so the end-to-end delay will decrease. When the traffic load is large enough (number of CBR connection pairs is larger than 60), the network interference and the probability of buffer overflow increase, so the network capability will close to saturation or over-saturated, which increases the end-to-end delay. This also can be found in Fig. 10. In Fig. 10, when the number of vehicles is less than 60, the decreasing of the packet delivery ratio is slight, so considering the increasing of the traffic load, the end-to-end delay decreases; similarly, when the number of CBR connection pairs is larger than 60, the decreasing of the packet delivery ratio is sharp, which contributes to the increasing of the end-to-end delay. Actually, the results shown in Fig. 11 also can be used to explain the conclusion in Fig. 10. Moreover, since the PRO routing algorithm takes the SINR and the packet queue length into account, so the performance of end-to-end delay is better and more stable than the other three routing algorithms. For instance, the end-to-end delay in PRO algorithm is about 1s less than that in SRPE algorithm; the variation of the end-to-end delay in PRO algorithm is less than 0.2s, which is about 1s in GPSR algorithm.

The network throughputs under different traffic load are shown in Fig. 12. Due to the excellent performance of packet delivery ratio and end-to-end delay, the network throughput of PRO routing algorithm is much better than that of the other three routing algorithms. With the increasing of the traffic load, the network throughputs of these four routing algorithm decrease. This is because the increasing of the network interference; moreover, the packet queue length is large when the network traffic load is heavy. But since the PRO routing algorithm takes the network interference and the packet queue length into account during the routing decision, so the decreasing of the network throughput in PRO algorithm is slight and the performance of network throughput is stable.

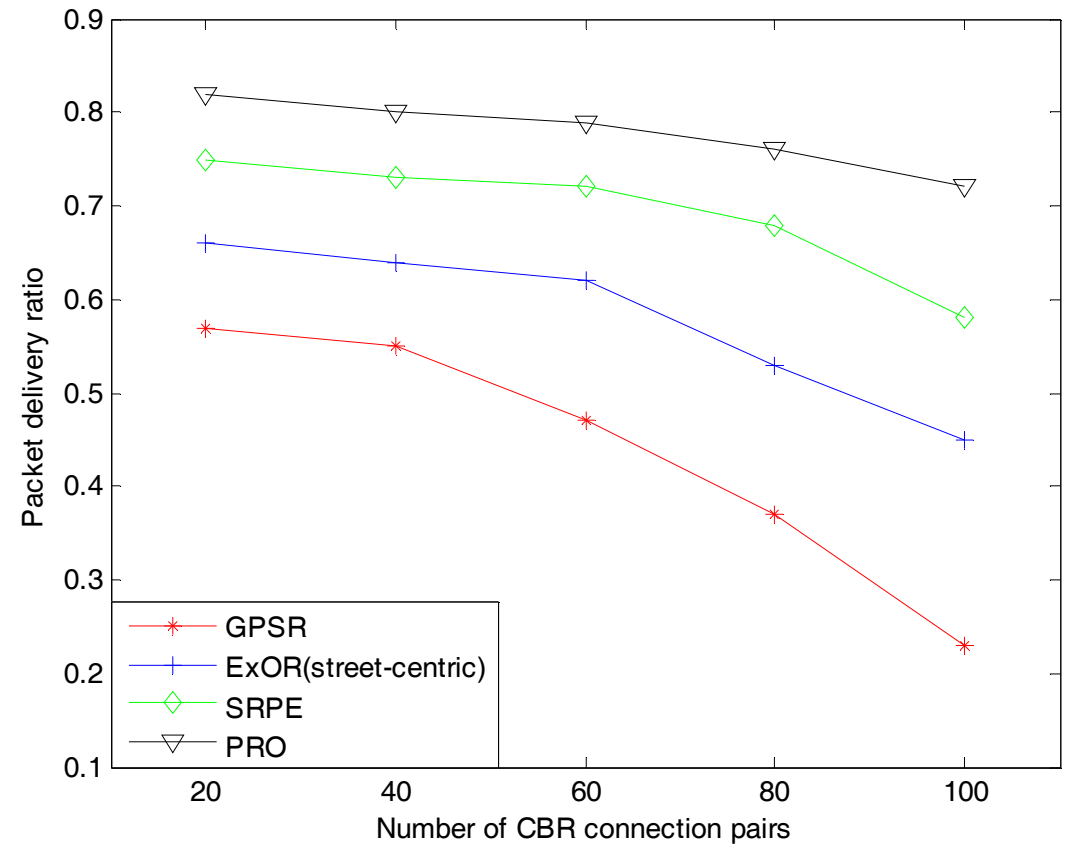

Fig. 10. Packet delivery ratio under different number of CBR connection pairs

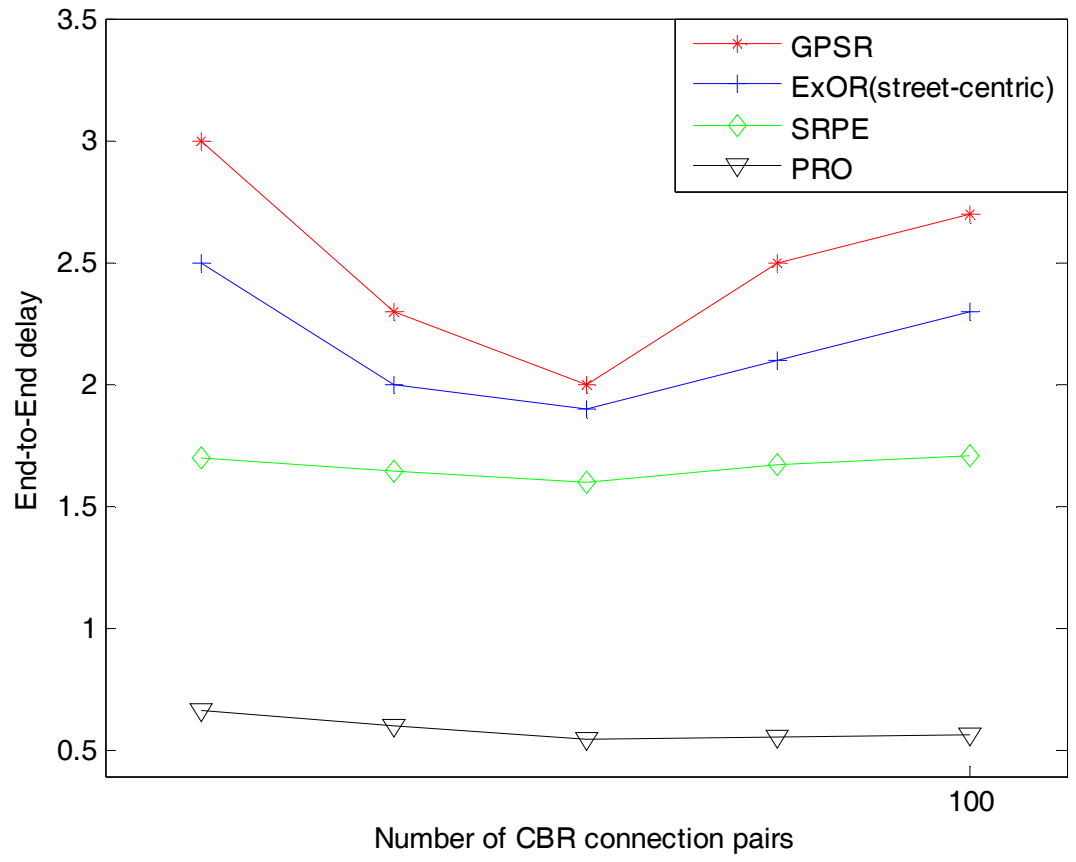

Fig. 11. End-to-End delay under different number of CBR connection pairs

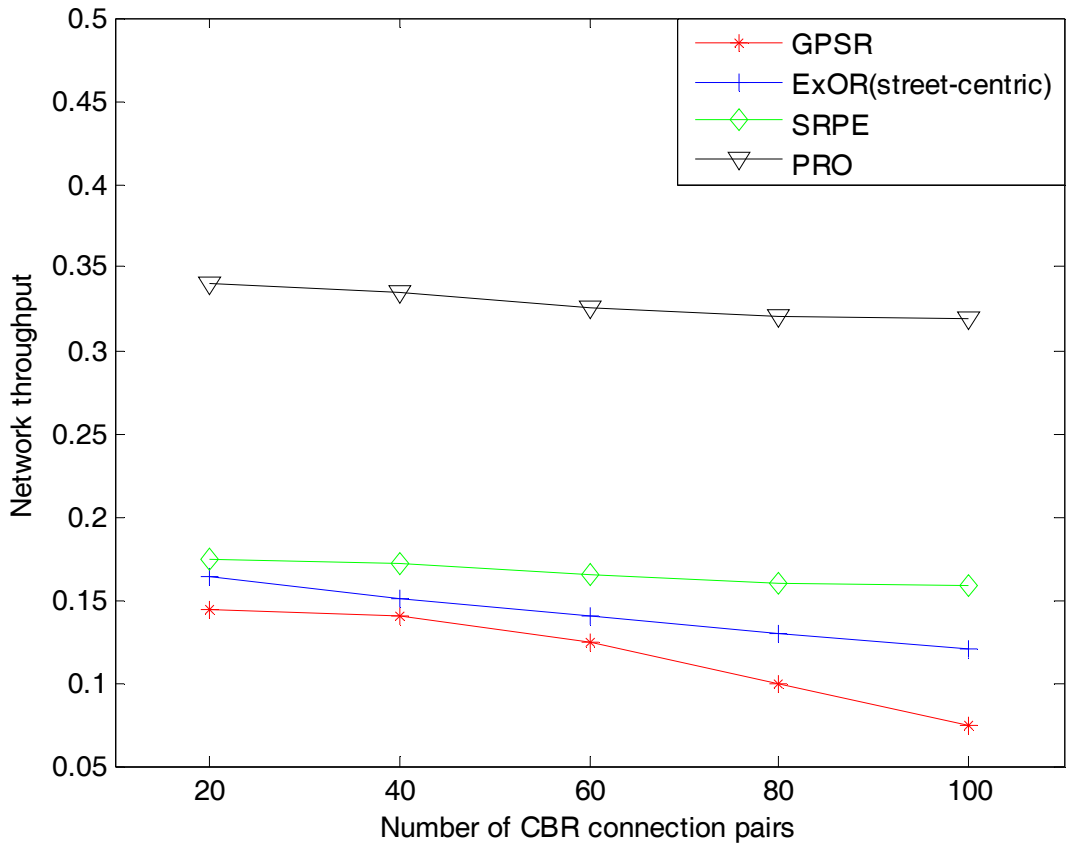

Fig. 12. Network throughput under different number of CBR connection pairs

## VII. CONCLUSION

In this paper, we propose the probability prediction based reliable opportunistic routing for the VANETs. The PRO routing algorithm can predict the variation of SINR and packet queue length in the receiver, which are $Pr_{sr}^{SIR}(w)$ and $Pr_r^Q(\Delta t)$, respectively. The prediction results are used to determine the utility of each relaying vehicle in the candidate set. The calculation of the vehicle utility is weight based algorithm. The weights are the variances of $Pr_{sr}^{SIR}(w)$ and $Pr_r^Q(\Delta t)$. The relaying priority of each relaying vehicle is determined by the value of the utility. By these innovations, the PRO can achieve better routing performance (such as the packet delivery ratio, the end-to-end delay, and the network throughput) than the SRPE, ExOR (street-centric), and GPSR routing algorithms.

## Appendix A

If variable x follows a normal distribution with mean $\mu$ and variance $\sigma^2$, then the probability distribution function and the distribution function of x can be expressed as:

$$f_X(x) = \frac{1}{\sqrt{2\pi}\sigma} e^{-\frac{(x-\mu)^2}{2}} \tag{27}$$

$$F_X(x) = \frac{1}{\sqrt{2\pi}\sigma} \int_{-\infty}^{x} e^{-\frac{(x-\mu)^2}{2}} dx \tag{28}$$

where $f(x) = F'(x)$.

Therefore, if $y = x^{-\alpha}$, then the distribution function of y can be expressed as:

$$\begin{aligned} F_Y(y) &= P\{y \le Y\} = P\{x^{-\alpha} \le Y\} \\ &= P\left\{x \ge Y^{-\frac{1}{\alpha}}\right\} = 1 - F_X(Y^{-\frac{1}{\alpha}}) \end{aligned} \tag{29}$$

Since $f(x) = F'(x)$, so the probability distribution function of y can be gotten by calculating the derivation on (29), which can be expressed as:

$$f_Y(y) = \frac{1}{\alpha} y^{-\frac{1+\alpha}{\alpha}} f_X(y^{-\frac{1}{\alpha}}) \tag{30}$$

According to (27), the probability distribution function of y expressed in (30) can be calculated as:

$$f_Y(y) = y^{-\frac{1+\alpha}{\alpha}} \frac{1}{\sqrt{2\pi}\alpha} e^{-\frac{\left(y^{-\frac{1}{\alpha}}-\mu\right)^2}{2}} \tag{31}$$

## Appendix B

When $x \sim N(\mu, \alpha^2)$, for each $x^{-\alpha}$, the probability distribution function can be found in (31); moreover, the $x_i^{-\alpha}$ ( $i = 1,2,...,m$ ) are independent identically distributed. Based on the principle of the distribution of multidimensional random variables, the probability distribution function of $z = \sum_{i=1}^{m} x_i^{-\alpha}$ can be calculated. For more clearly, the $z_i$ ( $i = 1,2,...,m$ ) can be expressed as:

$$\begin{aligned} & z_1 = y_1 \\ & z_2 = z_1 + y_2 \\ & \vdots \\ & z_{m-1} = z_{m-2} + y_{m-1} \\ & z_m = z_{m-1} + y_m \end{aligned} \tag{32}$$

where $y_i = x_i^{-\alpha}$.

Therefore, according to the principle of the distribution of multidimensional random variables, the probability distribution function of $z_m = y_1 + y_2 + ... + y_m$ can be calculated as:

$$f_{z_m}(z_m) = \int_{-\infty}^{\infty} f_{z_{m-1}}(z_m - y_m) f_{y_m}(y_m) dy_m \tag{33}$$

where $z_m$ and $z_{m-1}$ can be found in (32). Similar to (33), the probability distribution function of $z_{m-1} = z_{m-2} + y_{m-1}$ can be

calculated as:

$$f_{z_{m-1}}(z_{m-1})=\int_{-\infty}^{\infty}f_{z_{m-2}}(z_{m-1}-a_{m-1})f_{a_{m-1}}(a_{m-1})da_{m-1} \qquad (34)$$

Substituting (34) into (33), and noticing that $z_{m-1}=z_m-y_m$, the (33) can be rewritten as:

$$f_{z_m}(z_m)=\int_{-\infty}^{\infty}\int_{-\infty}^{\infty}f_{z_{m-1}}(z_m-y_m-y_{m-1})f_{y_m}(y_m)f_{y_{m-1}}(y_{m-1})dy_m dy_{m-1} \quad (35)$$

Similarly to (35), the $f_{z_{m-i}}(z_{m-i})$ ( $i=1,2,...,m$ ) can be calculated. Then $f_{z_{m-i}}(z_{m-i})$ ( $i=1,2,...,m$ ) are iterative substituted into (33), by which the probability distribution function of $z=\sum_{i=1}^{m}x_i^{-\alpha}$ can be calculated as:

$$f_z(z)=\underbrace{\int_{-\infty}^{\infty}\cdots\int_{-\infty}^{\infty}}_{m}f_{z_1}(z_m-\sum_{i=2}^{m}y_i)\prod_{i=2}^{m}f_{y_i}(y_i)dy_2\cdots dy_m \qquad (36)$$

Moreover, substitute (31) into (36), and considering the probability distribution function of N is $p(N)$, then the (36) can be rewritten as:

$$f_z(N+z)=p(N)+\underbrace{\int_{-\infty}^{\infty}\cdots\int_{-\infty}^{\infty}}_{m}\left(z-\sum_{i=2}^{m}y_i\right)^{-\frac{1+\alpha}{\alpha}}\frac{1}{\sqrt{2\pi}\alpha}e^{-\frac{\left(\left(z-\sum_{i=2}^{m}y_i\right)^{-\frac{1}{\alpha}}-\mu\right)^2}{2}}$$

$$\cdot\prod_{i=2}^{m}y_i^{-\frac{1+\alpha}{\alpha}}\frac{1}{\sqrt{2\pi}\alpha}e^{-\frac{\left(y_i^{-\frac{1}{\alpha}}-\mu\right)^2}{2}}dy_2\cdots dy_m \qquad (37)$$

## Appendix C

When $x\sim N(\mu,\alpha^2)$, for each $x^{-\alpha}$, the probability distribution function can be found in (31); moreover, since the $x_i^{-\alpha}$ ( $i=1,2,...,m$ ) are independent identically distributed, so the probability distribution function $z=\sum_{i=1}^{m}x_i^{-\alpha}$ can be calculated by (36). Therefore, the probability distribution function of $w=y/z$ can be calculated.

Since the y and z are independent, according to the principle of the distribution of multidimensional random variables, the probability density function of w can be calculated as:

$$\begin{aligned}f_W(w)&=\int_{-\infty}^{\infty}|z|f(wz,z)dz=\int_{-\infty}^{\infty}|z|f_Z(z)\cdot f_Y(wz)dz\\&=\int_{-\infty}^{\infty}|z|(wz)^{-\frac{1+\alpha}{\alpha}}\frac{1}{\sqrt{2\pi}\alpha}e^{-\frac{\left((wz)^{-\frac{1}{\alpha}}-\mu\right)^2}{2}}\cdot f_Z(z)dz\end{aligned} \qquad (38)$$

where $f_Z(z)$ can be calculated by (37).

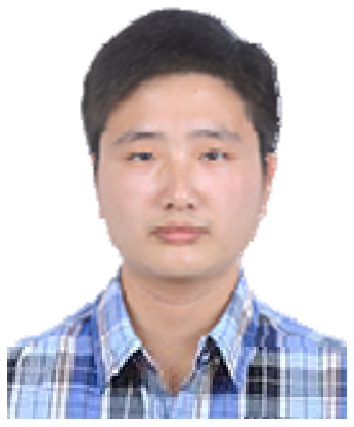

**Ning Li** received his B.S. degree and M.S. degree in Zhengzhou University, China, in 2010 and 2014, respectively. He is a Ph.D student in the Universidad Politecnica de Madrid research group Next-Generation Networks and Services. He has authored many journal papers (SCI indexed). His main research interests are within the area of mobile ad hoc network, network protocols, topology control, network optimization, and distributed computing. He is a student member of IEEE.

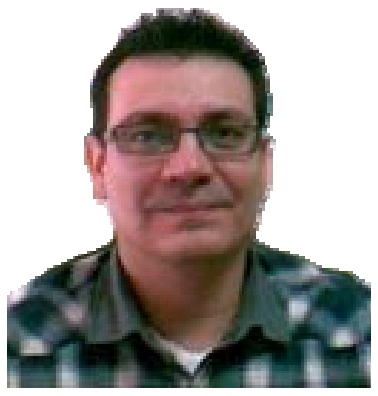

**José-Fernán Martínez** received his Ph.D. degree in Telematic Engineering from the Technical University of Madrid (UPM) Spain in 2001. He is Associate Professor at DIATEL (Department of Engineering and Telematic Architectures) of the same University. He is graduated in Electronic and Telecommunications Engineering at 1993 where he started R&D tasks. Since 1993 to 1996, he works as technical responsible in research projects at National Telecommunications Company in Colombia (TELECOM) and he was technical manager in his own company the S&H Ltda. His main interest areas and expertise are Ubiquitous Computing and Internet of Things; Smart CIties, Wireless Sensor & Actuators Networks (WSAN); Next-Generation Telematic Network and Services; Software Engineering and Architectures; Distributed Applications and intermediation platforms (middleware); High performance and fault tolerant systems. He has authored several national and international publications included in the Science Citation Index in his interest areas, and and he is technical reviser and chair of technical national and international events on Telematics, as well as member of different international and scientific committees. He has participated on several International and European Projects.

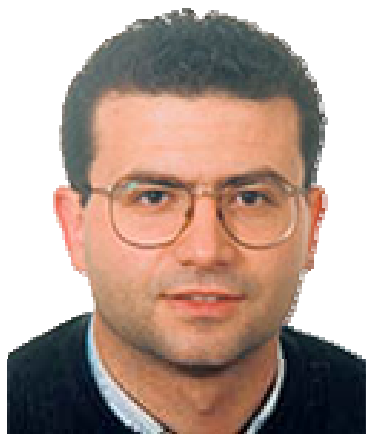

**Vicente Hernández Díaz** received his B.S. degree in Electronic Engineering from Universidad Politécnica de Madrid, Spain, in 1990 and his M.S. degree in Electronic Engineering from Universida de Alcala, Spain, in 2013. He was an assistant professor in Universidad Politécnica de Madrid (communication engineering) from 1990 to 1995. He has been working as a professor since 1995 in Universidad Politécnica de Madrid. He is a Ph.D. student in the research group Universidad Politécnica de Madrid research group Next-Generation Networks and Services in Universidad Politécnica de Madrid. He has authored or coauthored over 10 journal papers (SCI indexed) and coauthored three books. His main research interests are within the area of Internet of Things, architectural models, discovery of things and capabilities and self-adaptation.

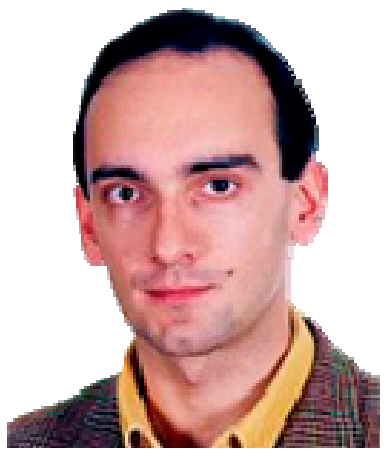

**Jose Antonio Sanchez Fernandez** received his degree in physics from the Universidad Complutense de Madrid, Spain, in 1989. From 1995 he has been mainly involved in management activities at the Universidad Politécnica de Madrid, where he is associate professor at the Department of Telematics and Electronics Engineering. His main areas of research include complex networks analysis, dynamics of telecommunication systems, and wireless sensor networks services and applications, which are the main topics related to his current Ph. D. thesis.